\documentclass[traditabstract]{aa}  
\usepackage{geometry}
\usepackage{xcolor}
\usepackage{colortbl}
\usepackage{tabularx}
\usepackage{graphicx} 

\usepackage{multirow}

\definecolor{lightgray}{rgb}{0.83, 0.83, 0.83}
\definecolor{lightblue}{rgb}{0.67, 0.84, 0.90}
\definecolor{lightgreen}{rgb}{0.56, 0.93, 0.56}

\usepackage{natbib}
\bibpunct{(}{)}{;}{a}{}{,} 

\usepackage{graphics}   
\usepackage{graphicx}   
\usepackage{epstopdf}
\usepackage{longtable}   
\usepackage{url}      
\usepackage{bm}        
\usepackage{soul}
\usepackage{comment}
\usepackage{threeparttable}
\usepackage{soul}
\usepackage[varg]{txfonts}
\usepackage{pdflscape}
\usepackage{supertabular}
\usepackage{hyperref}
\usepackage{xcolor}
\usepackage[normalem]{ulem}
\usepackage{natbib}
\usepackage{amsmath}
\usepackage{multirow}
\usepackage{color}
\definecolor{green}{rgb}{0.3,0.7,0.}
\definecolor{purple}{rgb}{0.77, 0.29, 0.55}

\hypersetup{
    bookmarks=true,         
    unicode=false,          
    colorlinks=true,       
    linkcolor=blue,          
    citecolor=blue,        
    filecolor=blue,      
    urlcolor=blue           
}

\begin{document}

\title{Fast-rotating massive Population~III stars as possible sources of extreme N-enrichment in high-redshift galaxies}
\titlerunning{Nitrogen enrichment by rotating massive stars}
\author{Devesh Nandal, Yves Sibony, and Sophie Tsiatsiou}
\authorrunning{N.S.T.}

\institute{D\'epartement d'Astronomie, Universit\'e de Gen\`eve, Chemin Pegasi 51, CH-1290 Versoix, Switzerland}

\date{}

\abstract{ 
We present an analysis of the chemical compositions in high-redshift galaxies, with a focus on the nitrogen-enhanced galaxies GN-z11 and CEERS-1019. We use stellar models of massive stars with initial masses ranging from 9 to 120\,$M_\odot$ across various metallicities to deduce the chemical abundances of stellar ejecta for a few light elements (H, He, C, N, O). Our study reveals insights into the chemical processes and elemental synthesis in the early universe. We find that Population III stars, particularly at initial fast equatorial rotation and sampled from a top-heavy initial mass function, as well as stars at $Z=10^{-5}$ with moderate rotation, align closely with observed abundance ratios in GN-z11 and CEERS-1019. These models demonstrate $\log{\rm (N/O)} = -0.38, \log{\rm (C/O)} =-0.22$ and $\log{\rm (O/H)} + 12 = 7.82$ at dilution factors of $f \sim 20~\text{--}~100$, indicating a good match with observational data. Models at higher metallicities do not match these observations, highlighting the unique role of Population III and extremely metal-poor stars in enhancing nitrogen abundance in high-redshift galaxies. Predictions for other abundance ratios, such as $\log{\rm(He/H)}$ ranging from -1.077 to -1.059 and $\log{\rm(^{12}C/^{13}C)}$ from 1.35 to 2.42, provide detailed benchmarks for future observational studies.}


\maketitle

\section{Introduction}

The observation of nitrogen-enhanced high-redshift (high-$z$) galaxies, particularly in GN-z11 \citep{Oesch2016} and CEERS-1019 \citep{Finkelstein2017}, stands as a testament to the intricacies involved in deciphering the cosmos. GN-z11, placed at an unprecedented redshift of $\mathrm{z} = 10.6$ \citep{Bunker2023}, showcases a N/O abundance ratio that is not only more than four times solar \citep{Cameron2023}, but also challenges previously held assumptions of such high-$z$ objects being nitrogen-poor. This ratio, which is elevated in comparison to galaxies at lower redshifts, and even higher than in some galaxies with super-solar metallicities, points towards peculiar chemical processes in the early universe \citep{Vincenzo2016,Berg2019}.

CEERS-1019's characteristics add further complexity to the narrative as highlighted by the works of \citet{Nakajima2023, Larson_2023}. Residing at a redshift of $z=8.678$, its emission lines, particularly the prominent N IV] $\lambda1486$, suggest advanced chemical processes for a galaxy of its era \citep{Tang2023}. Following a detailed analysis of UV spectra of galaxies at high redshifts, \citet{Isobe2023} found CEERS-1019 to exhibit extremely low C/N (${\rm (C/N)} \leq -1$) and high N/O (${\rm(N/O)} \geq 0.5$) ratios. \citet{Marques2023} found that CEERS-1019, similar to GN-z11, has a peculiar rest-frame UV spectrum which strongly hints to it being a N-emitter. They further provided accurate chemical abundances based on emission lines of H, C, N, O and Ne. The abnormal abundances observed therein beckon for a deeper investigation into the universe's nucleosynthesis and star formation processes. 

Although rare, other strong N-emitters in the high and low redshift universe have been identified in the last two decades. A few notable targets depicting high N/O include the Lynx arc \citep[$z=3.357$, $\log{\rm(N/O)}\sim-0.53$,][]{Fosbury2003,Villar2004}, sunburst \citep[$z=2.37$, $\log{\rm(N/O)}\sim-0.21$,][]{Rivera2019, Vanzella2022}, SMACS2031 \citep[$z=3.5$, $\log{\rm(N/O)}\sim-0.66$,][]{Patricio2016}, as well as Mrk 996 \citep[$z=0.00540$, $\log{\rm(N/O)}\sim-0.14$,][]{Mingozzi2022} in the low redshift Universe. It is increasingly apparent that massive, very massive, and supermassive stars (MSs, VMSs, and SMSs respectively) could play an integral role in the chemistry of these galaxies \citep{Meynet2006, Ohkubo2009, Woods2020}. Their potential nucleosynthetic contribution intertwines stellar evolution with the chemistry of high-redshift galaxies, but the exact mechanisms remain largely unexplored \citep[][]{DenissenkovHartwick2014, BastianLardo2018, Gieles2018a, Vink2023,Nandal2024}. Furthermore, their potential role in the formation of central massive black holes ties stellar evolution with galactic morphology and development \citep{Hosokawa2010, Natarajan_2023}.

In light of these observations, numerous studies have attempted to dissect the peculiarities of early universe galaxies. \citet{Kobayashi2023} emphasised a dual-burst model of rapid chemical enrichment, suggesting that Wolf-Rayet (WR) stars are crucial contributors during the secondary burst phase. \citet{Prantzos2017} and \citet{Gratton2019} provided a deeper understanding of the chemical peculiarities commonly found in globular clusters from the early Universe, which resonate with patterns observed in galaxies like GN-z11. Building on these perspectives, both \citet{Charbonnel2023} and \citet{Nagele2023} probed the profound impact that metal-enriched stars, especially those within the mass range of $10^3 - 10^5\,M_\odot$, can have on the chemical properties of early galaxies. {\citet{Isobe2023} have looked at the chemical abundances of 70 galaxies and explored the N/O, C/O, S/O, Ar/O and Ne/O ratios in regards to WR, SMS and tidal disruption events. 

Expanding this horizon further, various works have delved into different channels of stellar evolution. The role of fast-rotating massive stars, for instance, has been meticulously analysed by \citet{Meynet2006} and \citet{Choplin2018}. Their insights suggest that the rapid rotation of these stellar giants could significantly influence their nucleosynthetic outputs, a hypothesis further supported by chemical evolution models of the Galaxy from \citet{Chiappini2003, Chiappini2006, Chiappini2013}. The potential contribution of pair-instability supernovae (PISN), as explored by \citet{Heger2002,Takahashi2018}, also promises a deeper understanding of the stellar processes in play during the early epochs of the universe.

In response to this rich tapestry of research, our study embarks on a comprehensive exploration into stellar evolution. We compute the stellar yields of massive stars, ranging from 9 to 120\,$M_\odot$, at metallicities of 0 (Pop~III), $10^{-5}$ (Extremely metal-poor (EMP)), $4\times10^{-4}$ (IZw18 galaxy), 0.002 (Small Magellanic Cloud (SMC)), 0.006 (Large Magellanic Cloud (LMC)) and 0.014 (solar). We obtain yields from both stellar winds during evolution and a final ejection event. Due to uncertainties in determining the ejected mass at this final event \citep{Sukhbold2014, Sukhbold2018, Sukhbold2020}, we take an agnostic approach with two extreme scenarios: one where all mass above the remnant is lost, the other where all mass above the CO core is ejected. Using two different initial mass functions (IMF) for the population synthesis, we aim to predict the overall enrichment in terms of N/O, C/O, O/H, He/H, and $^{12}$C/$^{13}$C. Our results are then juxtaposed with the observed data from GN-z11 and CEERS-1019, corresponding to low metallicities of $\log{\rm(O/H)}+12=\mathrm{7.84}_{-0.05}^{+0.06}$ \citep{Senchyna2023} and $\log{\rm(O/H)}+12=7.70\pm0.18$ \citep{Marques2023}, respectively, seeking to offer a fresh perspective on the questions of elemental abundance patterns in high-redshift galaxies. The goal of this paper is to explore the various chemical abundances expected to be ejected from massive stars in the first galaxies. This paper is set out as follows: Sect.~\ref{Sec:Methods} describes the stellar evolution models and methods used to obtain the remnant mass of models and enrich the interstellar medium (ISM). Section~\ref{Sec:Results} first explores the mechanisms invoked for the transport of chemical species, then presents the ejected abundances by individual stars and the abundance ratios in the enriched ISM. Section~\ref{Sec:Discussion} discusses some caveats and compares our results to previous studies. Section~\ref{Sec:Conclusion} summarises our findings and provides prospects for future work. 

\section{Methods} \label{Sec:Methods}

Stellar models presented in this work have been computed using the Geneva stellar evolution code ({\sc Genec}) and the reader can find more details about the models in the corresponding papers {\citep[][]{Ekstrom2012, Georgy2013a, Groh2019, Eggenberger2021, Murphy2021a, Tsiatsiou2024, Sibony2024}. The models have masses of 9, 20, 60, 85, and 120\,$M_\odot$, with metallicities of $Z=0$ (Pop~III), $Z=10^{-5}$ (EMP), $Z=4\times10^{-4}$ (IZw18), $Z=0.002$ (SMC), $Z=0.006$ (LMC), and $Z=0.014$ (solar), at different initial surface velocities. The Pop~III models are fast-rotating with $\upsilon_{\mathrm{ini}}/\upsilon_{\mathrm{crit}} = 0.7$, whereas all other models are computed at a moderate rotation of $\upsilon_{\mathrm{ini}}/\upsilon_{\mathrm{crit}} = 0.4$. $\upsilon_{\mathrm{ini}}$ is the equatorial velocity on the zero age main sequence (ZAMS) and $\upsilon_{\rm crit}$ is the critical rotation velocity\footnote{The critical rotation velocity is defined as the velocity for which the gravitational acceleration is counterbalanced by the centrifugal force at the equator, following the $\Omega\Gamma$ limit by \citet{OGlimit2000}.}, computed as in \citet{Georgy2013a} and \citet{Tsiatsiou2024}.

For the non-Pop~III models, we use a metallicity-dependent mass-loss rate $\dot{M}$ during the main sequence, supergiant and Wolf-Rayet, with $\dot{M} = (Z/Z_\odot)^\alpha \dot{M}_\odot$, with $\alpha=0.85$ \citep{Vink2001}, $\alpha=0.5$ \citep{DeJager1988}, or $\alpha=0.66$ \citep{Eldridge2006}, depending on the case. For red supergiant stars, the mass-loss rate depends only on stellar luminosity and not on metallicity. \citep[more details can be found in][]{Ekstrom2012,Georgy2013b}. For the Pop~III models, we do not take into account any wind mass loss \citep{Murphy2021a}, but the stars can still lose some (almost negligible) amount of mass \citep[][]{Murphy2021a, Tsiatsiou2024}, due to the mechanical mass loss. This process removes mass from the outer layers of the star when the surface rotation reaches the critical limit, thus staying subcritical.

The models do not take into account the effects of magnetic fields. Moreover, they are based on the shellular rotation theory by \citet{Zahn1992} and are considered to have a near uniform rotation along an isobar. The transport of chemical species is governed by a purely diffusive equation. The Pop~III models (except for the 120\,$M_\odot$ one, see Sect.~\ref{Sec:caveats}) are computed until at least the end of the core He-burning phase and the other metallicities reach at least the end of the core C-burning phase. Numerical tests indicate the chemical composition of the ejecta in these models does not vary in a significant manner between the end of core He-burning and the end of core C-burning phase.

The stellar yields from our massive star models do not take explosive nucleosynthesis during the supernova into account. We base this assumption on the findings by \citet{Thielemann1990} where it was shown that for models undergoing explosive nucleosynthesis, the amount of $^{12}$C, $^{14}$N and $^{16}$O ejected is unaffected.

\subsection{Integrated abundance ratios} \label{Sec:ratio}

In this work, we study a scenario where stars enrich the chemical composition of their surrounding interstellar medium through mass loss. This might occur through a series of mass-loss episodes (such as stellar winds) prior to the star's death, or at the end of evolution through a supernova (SN) or the formation of a planetary nebula (we denote both cases as `SN' in the following for simplicity, as it is the case for all stars except for the 9\,$M_\odot$ ones). So the total mass ejected by a star of initial mass $M_{\rm ini}$ at metallicity $Z$, $M_{\rm ej}(M_{\rm ini}, Z)$, is given by $M_{\rm ej}(M_{\rm ini}, Z) = M_{\rm winds}(M_{\rm ini}, Z) + M_{\rm SN}(M_{\rm ini}, Z)$. In this study, we make the assumption of instantaneous mixing of the ejecta in the ISM, which amounts to summing all the mass lost during the evolution to the mass ejected at the star's death, and diluting this total ejected mass into the ISM. We now give a brief overview of the steps we undertake in order to obtain the diluted abundances of individual elements ejected by a population of stars.

For a given element, we first obtain from our stellar models the mass that is ejected at each timestep by mass loss. We then integrate its remaining abundance between the remnant mass \citep[which we determine by using the results of][see more details in Sect.~\ref{Sec:remnants}]{Farmer2019,Patton2020} and the surface in the final computed timestep. Once we have the mass of each element ejected by stars of each initial mass, at each metallicity, we generate populations of stars following an IMF (see details in Sect.~\ref{Sec:IMF}) where each population corresponds to a given metallicity. This gives us the mass of each element ejected into the ISM by each stellar population. Finally, we dilute the ejecta into the ISM and compute the (N/O) and (C/O) abundance ratios in the ISM after enrichment. We provide more details about the computation of the remnant masses in Sect.~\ref{Sec:remnants}, Sect.~\ref{Sec:enrichment} presents the enrichment model, and Sect.~\ref{Sec:IMF} details the population synthesis model.

\subsubsection{Remnant masses} \label{Sec:remnants}

To obtain the chemical abundances at the last computed time of stellar evolution, one needs to set a lower bound for the integration. Everything below this mass-cut $M_{\rm cut}$ will be trapped in the remnant, the rest will be ejected. We compute this mass using the models from \citet{Farmer2019,Patton2020}. \citet{Farmer2019} evolved massive helium stars with varying physical ingredients to the end of their evolution to predict the location of the pair-instability mass gap. \citet{Patton2020} evolved naked CO cores with varying mass $M_{\rm CO} \in [2.5,10]\,M_\odot$ and carbon mass fraction $x(^{12}{\rm C})$, from the end of core He-burning to the end of stellar evolution. They then used the `Ertl criterion' \citep{Ertl2016,Ertl2020} to predict if the star will explode or implode. We distinguish five cases depending on the mass of the CO core at the end of He-burning (which we define as the region within which the mass fraction of helium $Y<10^{-2}$). If $M_{\rm CO}<1.4\,M_\odot$, the remnant will be a white dwarf (WD) with mass $M_{\rm rem} = M_{\rm CO}$. If $1.4\,M_\odot<M_{\rm CO}<2.5\,M_\odot$, the remnant will be a neutron star (NS) with mass $M_{\rm rem}=1.4\,M_\odot$. If $2.5\,M_\odot<M_{\rm CO}<10\,M_\odot$, we get from our model the central carbon mass fraction at the end of He-burning $x(^{12}{\rm C})_c$ and look at Fig.~3 from \citet{Patton2020} to find if the star will explode and leave a NS, or implode into a black hole (BH). If it explodes, the mass of the NS will be $M_{\rm rem}=M_4$, where $M_4$ is the mass coordinate inside the star at which the entropy per baryon is equal to $4k_b$. We use Table~1 of \citet{Patton2020}, which gives values of $M_4$ for a range of $(M_{\rm CO},x(^{12}{\rm C})_c$ at the end of core He-burning. If it implodes we set the mass of the resulting BH to $M_{\rm rem}=\min{(M_{\rm CO}+4\,M_{\odot}, M_{\rm tot})}$, where $M_{\rm tot}$ is the total mass of the star at the final timestep, following the analytic fit provided by \citet[see their Appendix A]{Farmer2019}. If $10\,M_\odot<M_{\rm CO}<38\,M_\odot$, the star will implode, and we set the resulting mass to $M_{\rm rem}=\min{(M_{\rm CO}+4\,M_\odot, M_{\rm tot})}$. Finally, if $M_{\rm CO}>38\,M_\odot$, we use Table~1 from \citet{Farmer2019} to determine the fate of the star (by finding in their table the star with the closest CO core mass), as well as the remnant mass. In the case of a PISN, the remnant mass is 0 and all the star's matter is ejected.

We consider two scenarios: one where we use the formalism described in the previous paragraph, using $M_{\rm cut} = M_{\rm rem}$ (henceforth `above remnant'), and a much simpler one where we eject all matter above the mass coordinate of the CO core, with $M_{\rm cut} = M_{\rm CO}$ (henceforth `above CO core').

\subsubsection{Enrichment model} \label{Sec:enrichment}
\begin{table}[h]
    \centering
    \caption{Initial abundances (in mass fraction) of the isotopes of interest for the different metallicities we explore.}
    \resizebox{\columnwidth}{!}{
    \begin{tabular}{|c|cccccc|}
        \hline\hline
        Z & X$_{^1 \rm H}$ & X$_{^4 \rm He}$ & X$_{^{12} \rm C}$ & X$_{^{13} \rm C}$ & X$_{^{14} \rm N}$ & X$_{^{16} \rm O}$ \\
        \hline
        & \multicolumn{6}{c|}{Initial mass fraction}\\
        \hline
        0 & 0.7516 & 0.2484 & 0 & 0 & 0 & 0 \\ 
        $10^{-5}$ & 0.7516 & 0.2484 & 1.3e-06 & 4.3e-09 & 1.0e-07 & 6.8e-06 \\ 
        $4\times10^{-4}$ & 0.7507 & 0.2489 & 6.52e-05 & 7.92e-07 & 1.88e-05 & 1.63e-04 \\ 
        0.002 & 0.7471 & 0.2509 & 3.26e-04 & 3.96e-06 & 9.41e-05 & 8.17e-04 \\ 
        0.006 & 0.7381 & 0.2559 & 9.78e-04 & 1.19e-05 & 2.82e-04 & 2.45e-03 \\ 
        0.014 & 0.7200 & 0.2660 & 2.3e-03 & 2.8e-05 & 6.6e-04 & 5.7e-03 \\ 
        \hline
    \end{tabular}
    }
    \label{tab:initial_compositions}
\end{table}

The total mass of element $i$ ejected by a star of initial mass $M_{\rm ini}$ at metallicity $Z$, $M_{{\rm ej},i}(M_{\rm ini}, Z)$, is given by:
\begin{equation}
    M_{{\rm ej},i}(M_{\rm ini}, Z) = \int_{0}^{t_f}{M_{{\rm winds}}(t) x_{i,s}(t) dt} + \int_{M_{\rm cut}}^{M_{\rm tot}}{x_{i,t_f}(M_r) dM_r},
\end{equation}
where $t_f$ is the final age reached by the star, $M_{{\rm winds}}(t)$ the total mass ejected by winds at time $t$, $x_{i,s}(t)$ the surface mass fraction of element $i$ at time $t$, and $x_{i,t_f}(M_r)$ the mass fraction of element $i$ at the mass coordinate $M_r$ in the final computed timestep.

The mass fraction of element $i$ in the ejecta of the star, $X_{{\rm ej},i}(M_{\rm ini}, Z)$, is the ratio $M_{{\rm ej},i}(M_{\rm ini}, Z)/M_{\rm ej}(M_{\rm ini}, Z)$. For a population of stars with masses between $m_{\rm min}=9\,M_\odot$ and $m_{\rm max}=120\,M_\odot$, the mass fraction of element $i$ in the total ejecta, $X_{{\rm ej},i}(Z)$ is:
\begin{equation}
    X_{{\rm ej},i}(Z) = \frac{\sum\limits_{M_{\rm ini}} n(M_{\rm ini}) * M_{{\rm ej},i}(M_{\rm ini},Z)}{\sum\limits_{M_{\rm ini}} n(M_{\rm ini}) * M_{\rm ej}(M_{\rm ini}, Z)},
\end{equation}
where $n(M_{\rm ini})$ is the number of stars in the population with mass $M_{\rm ini}$ and depends only on our choice of an IMF (see Sect.~\ref{Sec:IMF}).

The ejected mass interacts with the ISM, leading to a mixture of the stars' ejecta and the ISM. The mass of the ISM involved in this interaction is $M_{\rm ISM}$. We introduce a dilution factor $f(Z)$, which is the ratio $M_{\rm ISM}/M_{\rm ej}(Z)$, where $M_{\rm ej}(Z)$ is the total mass ejected by the population at metallicity $Z$. Dilution factor is a free parameter and we allow it to vary in a range of values between 1 and 1000, in consistency with previously published works of \citet{Nagele2023} and  \citet{Charbonnel2023}.

Using this parameter, we estimate the final composition of the chemical mixture of the star's ejecta and the ISM. We define $X_{i,m}(Z)$ as the mass fraction of element $i$ in this mixture and it is calculated using the principle of mass conservation:

\begin{equation}
X_{i,m}(Z) * (M_{\rm ej}(Z) + M_{\rm ISM}) = M_{{\rm ej},i}(Z) + X_{i,0}(Z)* M_{\rm ISM},
\end{equation}
where $X_{i,0}(Z)$ is the mass fraction in the ISM prior to any enrichment. We provide in Table~\ref{tab:initial_compositions} the initial chemical compositions of the ISM for the six metallicities that we investigate in this paper. These are also the initial compositions of the stars. Indeed, stars form from the ISM gas, and the composition of the ISM does not change until the first elements are ejected from stars\footnote{The underlying assumption is that all stars form at the same time (so that no star can enrich the material from which other stars in the same population form).}.

By substituting the dilution factor $f(Z)$ into the equation, we can simplify it to:

\begin{equation}
X_{i,m}(Z)*(1+f(Z)) = X_{{\rm ej},i}(Z) + f(Z)*X_{i,0}(Z),
\end{equation}
or, isolating $X_{i,m}(Z)$:
\begin{equation}
X_{i,m}(Z) = \frac{X_{{\rm ej},i}(Z) + f(Z)*X_{i,0}(Z)}{1 + f(Z)}.
\end{equation}

Finally, we compute the logarithm abundance ratios (in number fraction) in the diluted ISM between isotopes $i$ and $j$, $\log{\rm(i/j)}(Z)$:
\begin{equation}
    \log{\rm(i/j)}(Z) = \log{\left(\frac{X_{i,m}(Z)}{X_{j,m}(Z)}\frac{A_j}{A_i}\right)},
\end{equation}
where $A_i$ and $A_j$ are the atomic masses of $i$ and $j$. In the rest of this paper, all the abundance ratios that we provide are number ratios.

\subsubsection{Impact of initial mass function} \label{Sec:IMF}

We generate populations of massive stars from power-law IMFs $dN/dM_{\rm ini}\sim M_{\rm ini}^{\alpha}$. We consider either a Salpeter ($\alpha=-2.35$) or a top-heavy ($\alpha=-1$) IMF, with stellar masses between $m_{\rm min}=9\,M_\odot$ and $m_{\rm max}=120\,M_\odot$. Since our stellar models at all metallicities have initial masses of 9, 20, 60, 85, and 120\,$M_\odot$, we divide the continuous mass distributions into bins centred on these masses. We only consider massive stars (with an initial mass $M_{\rm ini} > 9\,M_\odot$) to enrich the ISM.

In the end, we explore four scenarios for each metallicity: two ways of determining the mass-cut in stars (`above remnant' or `above CO core', see Sect.~\ref{Sec:remnants}) above which to eject material, and two different IMFs (Salpeter or top-heavy). We name them `Salpeter above remnant', `Salpeter above CO core', `Top-heavy above remnant', and `Top-heavy above CO core'.

    \begin{figure}
    \centering
    \includegraphics[width=0.45\textwidth]{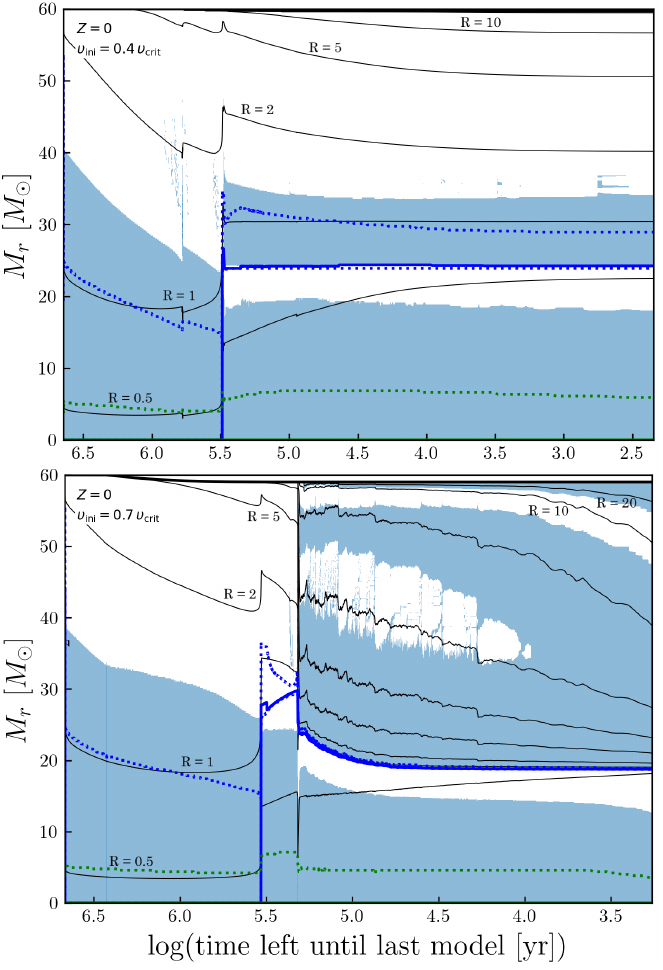}
    \caption{Kippenhahn diagrams displaying the structural evolution in lagrangian coordinates. Pop~III models are presented with 60\,$M_\odot$, and with an initial rotation of 40\% \citep[by][{\it top panel}]{Murphy2021a} and 70\% \citep[by][{\it bottom panel}]{Tsiatsiou2024} of the critical rotation velocity. The last evolutionary stage for both models is the end of the core He-burning phase. The blue regions are convective zones, while the white ones are radiative. The black lines are iso-radius lines (in units of solar radius $R_\odot$). The blue and green lines show the mass coordinates where hydrogen and helium are burning, respectively. The solid lines indicate the peak of the energy generation rate and the dashed ones delimit 10\% of the peak energy generation rate for each burning phase.}
    \label{fig:kippen}
    \end{figure}

    \begin{figure}[!h]
	\centering
	\includegraphics[width=9cm]{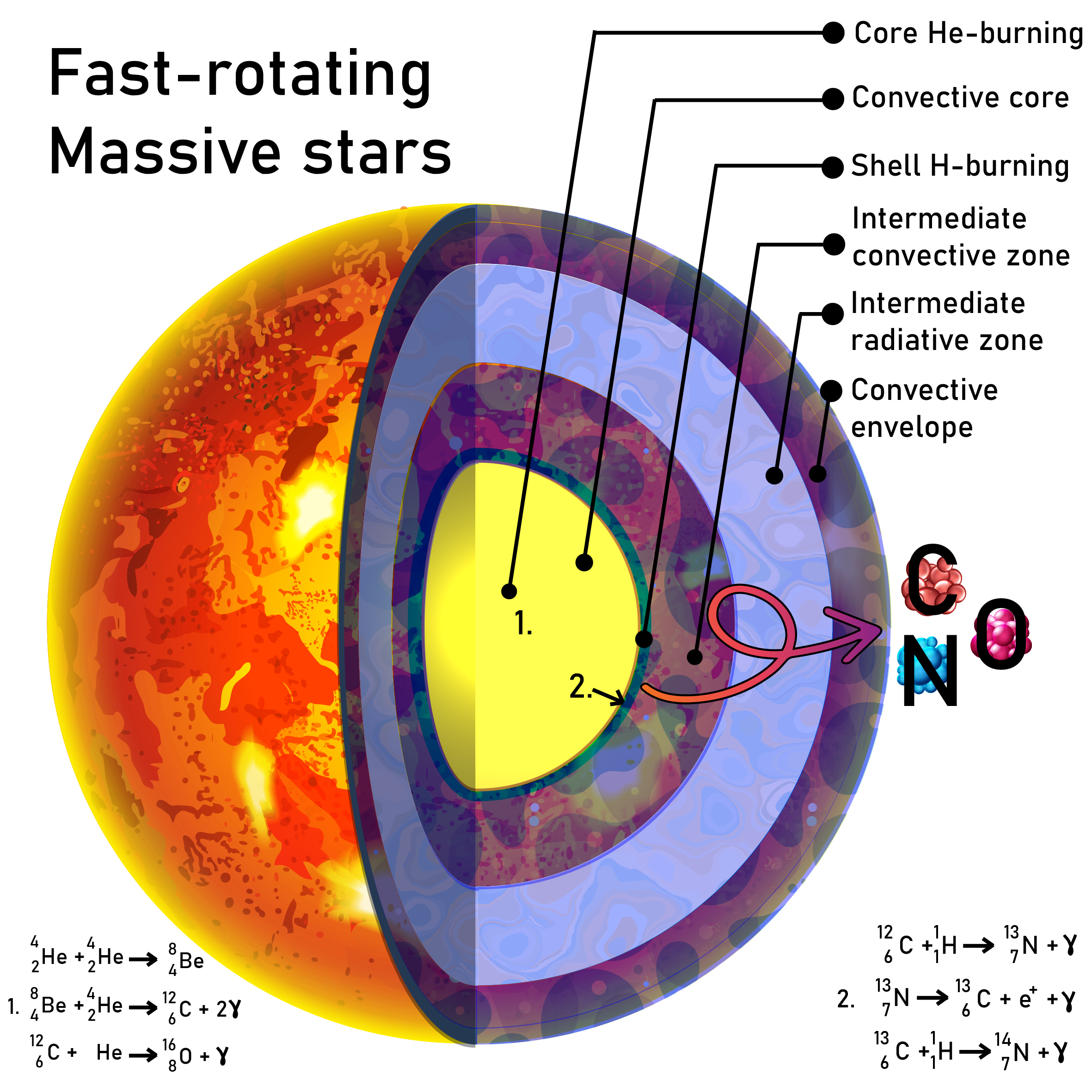}
	\caption{Illustration depicting the interior and the transport of chemical species in fast-rotating massive stars during core He-burning phase. Number 1 denotes the region where $^{12}$C and $^{16}$O are produced and number 2 shows the subsequent production of $^{14}$N. We also indicate the relevant nuclear reactions.}
	\label{fig:Illu}
    \end{figure} 

\section{Results} \label{Sec:Results}

\subsection{Mechanism of the transport of chemical species} \label{Sec:Transport}

Rotating massive star models undergo a drastic change in their structure after the end of the core H-burning phase. Here we compare the structure of two Pop~III massive stars with the same initial mass. Figure~\ref{fig:kippen} shows Kippenhahn diagrams of the 60\,$M_\odot$ Pop~III models with moderate \citep[top panel, $\upsilon_{\mathrm{ini}}/\upsilon_{\mathrm{crit}} = 0.4$, from][]{Murphy2021a} and fast \citep[bottom panel, $\upsilon_{\mathrm{ini}}/\upsilon_{\mathrm{crit}} = 0.7$, from][]{Tsiatsiou2024} rotation. Fig.~\ref{fig:Illu} illustrates the structure of fast rotating star as it commences core He-burning. Both models possess a fully convective core and a radiative envelope during core H-burning phase. As core He-burning commences, the fast-rotating model at 60\,$M_\odot$ (bottom panel of Fig.~\ref{fig:kippen}) has a convective core that extends towards the H-burning shell. The $^{12}$C produced via the triple-$\alpha$ reaction (denoted by 1 in Fig.~\ref{fig:Illu}) is transported to the H-burning shell via the extended convective core in the early stages of core helium burning. This $^{12}$C in the H-burning shell (denoted by 2 in Fig.~\ref{fig:Illu}) produces $^{13}$N which produces $^{13}$C. Finally, this $^{13}$C interacts with $^{1}$H to produce $^{14}$N; this is the CNO cycle. Additionally, the injection of $^{14}$N into the H-burning shell also boosts the energy production of the region and leads to the formation of an extended intermediate convective zone. The size of this convective zone depends on the rotation of the star as shown in Fig.~\ref{fig:kippen} and labelled as intermediate convective zone in Fig.~\ref{fig:Illu}. The moderately-rotating model has a smaller intermediate zone when compared to the fast-rotating model of the same mass. The consequence of the extended convective zone is the ability to transport the freshly produced $^{14}$N (denoted by 2 in Fig.~\ref{fig:Illu}) to the surface and this is most efficient in the case of the fast-rotating 60\,$M_\odot$ model. In this study, the fast-rotating model produces the strongest enrichment of $^{14}$N due to a higher injection of $^{12}$C in the H-burning shell via the extended convective core, when compared to the standard rotating model. Additionally, the same extended intermediate convective zones are also responsible for transporting $^{16}$O to the surface during the core He-burning phase.

Finally, as the energy production in the H-burning shell is boosted, the helium-rich core recedes and an intermediate radiative zone is formed between the He-burning core and H-burning shell. A detailed discussion on this transport mechanism is beyond the scope of this paper and will be described in more detail in \citet{Tsiatsiou2024}.

\subsection{Integrated abundances from individual stars} \label{Sec:Abund}

Here we explore the composition of material ejected by stars as a function of mass and metallicity. Figure~\ref{fig:Abunds_linscale} shows how the initial mass of each star is decomposed into what will be trapped in the remnant, and different elements that will be ejected during and at the end of its life. The top panel shows the composition assuming all layers above the remnant mass will be ejected by the final ejection event, and the bottom panel shows the case where all layers above the CO core will be ejected. Figure~\ref{fig:Abunds_linscale} does not show explicitly how much of the material is ejected by winds or by the supernova, but we discuss this aspect below. We find that hydrogen (depicted in black) is the dominant element in the 9\,$M_\odot$ stars, which eject about 35\% of their mass as hydrogen. In stars more massive than 20\,$M_\odot$, helium (grey) is the most abundant element in the ejecta, with its abundance varying based on metallicity. Carbon (red) ejection does not show a clear correlation with mass or metallicity. Nitrogen (green) production, significant in the 60\,$M_\odot$ Pop~III star (with 3.8\,$M_\odot$ or 6.2\% of its initial mass being ejected as nitrogen), is influenced by its fast rotation and convective structure. Oxygen (blue) is predominantly produced in later stages of stellar evolution and is the most abundant element in the ejecta of 120\,$M_\odot$ models at lower metallicities.

The total mass lost by all stars during their evolution due to mass-loss processes depends on mass and metallicity. For instance, a 60\,$M_\odot$ model at zero metallicity loses less than 1\% of its total mass whereas a 60\,$M_\odot$ model at solar metallicity loses around 70\% of its total mass to stellar winds. When comparing the wind-driven mass loss during the evolution for the 9 and 120\,$M_\odot$ models at $Z=10^{-5}$, we find that the 9\,$M_\odot$ model loses only 0.7\% of its total mass and the 120\,$M_\odot$ model around 28\%. At solar metallicity, the 9\,$M_\odot$ model loses 5\% and the 120\,$M_\odot$ 84\% of their initial mass to winds.

Comparing the top and bottom panels of Fig.~\ref{fig:Abunds_linscale}, the most striking differences appear for the 120\,$M_\odot$ models at $Z=0$ and $Z=10^{-5}$. Indeed, since we predict a PISN for these stars, the remnant mass will be 0. This ejects matter from layers deep within the stars which are rich in carbon and oxygen. For other stars, the discrepancy between the two ejection scenarios follows the opposite pattern, with more carbon and oxygen ejected above the CO core than above the remnant. This can be seen for the 20, 60, and 85\,$M_\odot$ Pop~III and EMP models. Indeed, we predict that these will form black holes with mass $M_{\rm BH}=M_{\rm CO}+4\,M_\odot$, and the layers that make up the 4\,$M_\odot$ above the CO core are rich in carbon and oxygen, explaining why the ejected material is richer in these elements when ejecting all matter above the CO core rather than above the remnant. The 9\,$M_\odot$ models present no difference between the ejection scenarios no matter the metallicity because we predict that they will become white dwarfs, with $M_{\rm WD}=M_{\rm CO}$. In their case, the final ejection event is not a supernova but rather a planetary nebula, nevertheless we consider that all mass above $M_{\rm WD}$ is ejected into and instantaneously mixed with the ISM.

\begin{figure*}[h]
    \centering
    \includegraphics{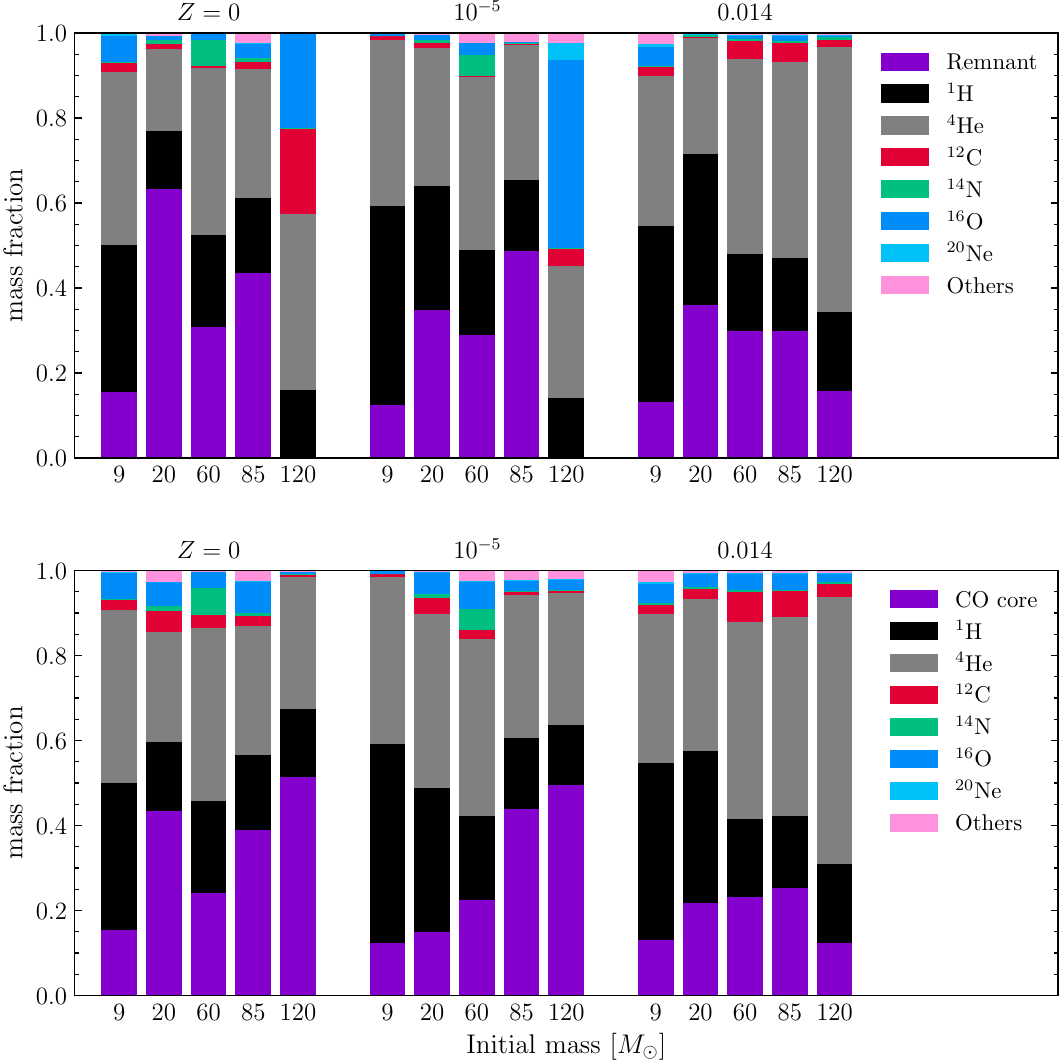}
    \caption{Partition of the initial mass of stars into remnant (or CO core) and ejecta (winds + final ejection event) for all the models we use. \textit{Top panel:} final ejection above the remnant mass. \textit{Bottom panel:} final ejection above the CO core mass.}
    \label{fig:Abunds_linscale}
\end{figure*}

\subsection{Abundance ratios in the diluted ejecta of stellar populations} \label{Sec:dilution}

\subsubsection{N/O and C/O abundance ratios} \label{Sec:ratios_obs}

\begin{figure*}[!]
    \centering\includegraphics{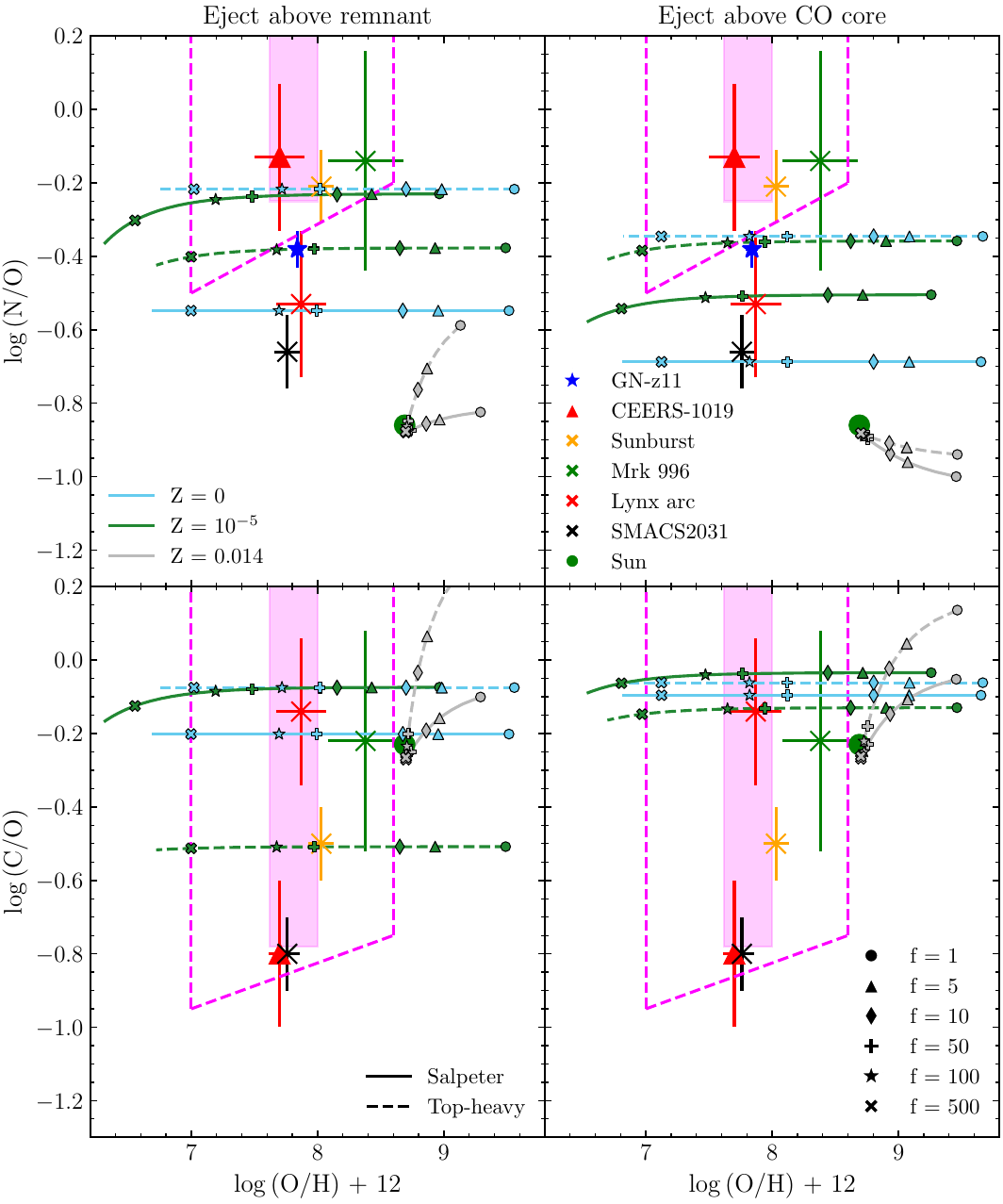}
    \caption{Abundance ratios in the diluted ISM with varying dilution factor $f=1 - 1000$. \textit{Top row:} $\log{\rm(N/O)}$ vs $\log{\rm(O/H)}+12$. \textit{Bottom row:} $\log{\rm(C/O)}$ vs $\log{\rm(O/H)}+12$. \textit{Left column:} scenarios where the stars eject everything above the remnant mass. \textit{Right column:} scenarios where the stars eject everything above the CO core. 
    In each panel, we show the results for three of the six metallicities: $Z=0$ (Pop~III, cyan), $Z=10^{-5}$ (EMP, green), and $Z=0.014$ (solar, grey). Solid lines represent populations sampled from a Salpeter IMF and dashed lines from a top-heavy IMF. The markers on these lines represent specific values for $f = 1 - 500$. The other markers (as indicated on the \textit{top right panel}), pink dashed lines and pink-shaded regions show results and lower bounds from observations, taken from \citet{Marques2023}. These results, as well as those for the remaining three metallicities not shown on this figure, can be found in tabular form in Tables~\ref{tab:all_ratios}~and~\ref{tab:obs_ratios}.}
    \label{fig:ratios}
\end{figure*}

\begin{table*}[h!]
    \centering
    \caption{Constrained values for $f$ and corresponding predicted abundance ratios for all metallicities in each scenario.}
    \resizebox{\textwidth}{!}{
    \begin{tabular}{|c|ccccccc|}
    \hline\hline
    Scenario & Z & $f$ & $\log{\rm(O/H)}+12$ & $\log{\rm(N/O)}$ & $\log{\rm(C/O)}$ & $\log{\rm(He/H)}$ & $\log{(^{12}{\rm C}/^{13}{\rm C})}$\\
    \hline
\multirow{6}{*}{\shortstack{Salpeter\\above remnant}} & 0 & 50 - 118 & 7.99 -- 7.62 & -0.55 & -0.20 & -1.07 -- -1.08 & 2.18 \\ 
& $10^{-5}$ & 15 - 36 & 7.99 -- 7.62 & -0.23 -- -0.24 & -0.08 & -1.05 -- -1.07 & 1.42 \\ 
& $4\times10^{-4}$ & 15 - 46 & 7.99 -- 7.62 & -1.57 -- -1.30 & -0.17 -- -0.19 & -1.05 -- -1.07 & 2.90 -- 2.52 \\ 
& 0.002 & 48 - 1000 & 8.00 -- 7.84 & -0.98 -- -0.89 & -0.23 -- -0.27 & -1.07 -- -1.08 & 2.15 -- 1.96 \\ 
& 0.006 & -- & 8.70 -- 8.32 & -0.55 -- -0.88 & 0.08 -- -0.27 & -0.86 -- -1.06 & 2.41 -- 1.95 \\ 
& 0.014 & -- & 9.29 -- 8.70 & -0.82 -- -0.88 & -0.10 -- -0.27 & -0.83 -- -1.03 & 2.37 -- 1.95 \\ \hline 
\multirow{6}{*}{\shortstack{Salpeter\\above CO core}} & 0 & 67 - 159 & 7.99 -- 7.62 & -0.69 & -0.10 & -1.07 -- -1.08 & 2.42 \\ 
& $10^{-5}$ & 29 - 70 & 8.00 -- 7.62 & -0.51 & -0.04 & -1.06 -- -1.08 & 1.68 \\ 
& $4\times10^{-4}$ & 14 - 41 & 7.97 -- 7.62 & -1.53 -- -1.29 & 0.01 -- -0.03 & -1.05 -- -1.07 & 3.06 -- 2.67 \\ 
& 0.002 & 78 - 1000 & 8.00 -- 7.85 & -1.01 -- -0.89 & -0.20 -- -0.26 & -1.07 -- -1.08 & 2.19 -- 1.98 \\ 
& 0.006 & -- & 9.20 -- 8.32 & -1.05 -- -0.88 & -0.04 -- -0.27 & -0.84 -- -1.06 & 2.80 -- 1.96 \\ 
& 0.014 & -- & 9.46 -- 8.70 & -1.00 -- -0.88 & -0.05 -- -0.27 & -0.81 -- -1.03 & 2.59 -- 1.96 \\ \hline 
\multirow{6}{*}{\shortstack{Top-heavy\\above remnant}} & 0 & 53 - 125 & 7.99 -- 7.62 & -0.22 & -0.07 & -1.07 -- -1.08 & 1.71 \\ 
& $10^{-5}$ & 47 - 114 & 8.00 -- 7.62 & -0.38 & -0.51 & -1.07 -- -1.08 & 1.35 \\ 
& $4\times10^{-4}$ & 45 - 139 & 8.00 -- 7.62 & -1.67 -- -1.34 & -0.58 -- -0.50 & -1.07 -- -1.08 & 2.49 -- 2.21 \\ 
& 0.002 & 187 - 1000 & 8.00 -- 7.87 & -1.03 -- -0.91 & -0.37 -- -0.30 & -1.07 -- -1.08 & 2.02 -- 1.96 \\ 
& 0.006 & -- & 8.92 -- 8.32 & -0.67 -- -0.88 & 0.26 -- -0.27 & -0.79 -- -1.06 & 2.70 -- 1.96 \\ 
& 0.014 & -- & 9.13 -- 8.70 & -0.59 -- -0.88 & 0.24 -- -0.27 & -0.75 -- -1.03 & 2.62 -- 1.96 \\ \hline 
\multirow{6}{*}{\shortstack{Top-heavy\\above CO core}} & 0 & 66 - 158 & 8.00 -- 7.62 & -0.35 & -0.06 & -1.07 -- -1.08 & 1.84 \\ 
& $10^{-5}$ & 44 - 106 & 8.00 -- 7.62 & -0.36 & -0.13 & -1.07 -- -1.08 & 1.64 -- 1.65 \\ 
& $4\times10^{-4}$ & 17 - 52 & 7.99 -- 7.62 & -1.55 -- -1.30 & -0.04 -- -0.08 & -1.04 -- -1.07 & 2.99 -- 2.61 \\ 
& 0.002 & 96 - 1000 & 8.00 -- 7.85 & -1.01 -- -0.90 & -0.23 -- -0.27 & -1.07 -- -1.08 & 2.16 -- 1.98 \\ 
& 0.006 & -- & 9.42 -- 8.32 & -1.18 -- -0.88 & 0.03 -- -0.27 & -0.78 -- -1.06 & 2.98 -- 1.97 \\ 
& 0.014 & -- & 9.47 -- 8.70 & -0.94 -- -0.88 & 0.14 -- -0.27 & -0.74 -- -1.03 & 2.86 -- 1.96 \\ \hline 
    \end{tabular}
    }
    \label{tab:all_ratios}
\end{table*}

\begin{table}[h!]
    \centering
    \caption{ Observational data shown in Fig.~\ref{fig:ratios}, obtained from Fig.~6 of \citet{Marques2023}.}
    \begin{threeparttable}
    \begin{tabular}{|c|ccc|}
    \hline\hline
    Galaxy & $\log{({\rm O}/{\rm H})}+12$ & $\log{({\rm N}/{\rm O})}$ & $\log{({\rm C}/{\rm O})}$ \\
    \hline
    GN-z11\tnote{$\dagger$} & $7.84^{+0.06}_{-0.05}$ & $-0.38^{+0.05}_{-0.04}$ &  --\\
    GN-z11\tnote{$\ddagger$} &  $7.82^{+0.18}_{-0.20}$ & $>-0.25$ & $>-0.78$ \\
    CEERS-1019 & $7.7\pm0.2$ & $-0.13\pm0.2$ & $-0.80\pm0.09$ \\
    Sunburst & $8.03\pm0.1$ & $-0.21\pm0.1$ & $-0.5\pm0.1$ \\
    Mrk 996 & $8.38\pm0.3$ & $-0.14\pm0.3$ & $-0.22\pm0.3$ \\
    Lynx arc & $7.87\pm0.2$ & $-0.53\pm0.2$ & $-0.14\pm0.2$ \\
    SMACS2031 & $7.76\pm0.1$ & $-0.66\pm0.1$ & $-0.8\pm0.1$ \\
    Sun & 8.69 & -0.86 & -0.23 \\
    \hline
    \end{tabular}
    \begin{tablenotes}
        \item [$\dagger$] Data from \citet{Senchyna2023}.
        \item [$\ddagger$] Data from \citet{Cameron2023}.
    \end{tablenotes}
    \end{threeparttable}
    \label{tab:obs_ratios}
\end{table}

Figure~\ref{fig:ratios} shows the $\log{\rm(N/O)}$ (top row) and $\log{\rm(C/O)}$ (bottom row) ratios as functions of $\log{\rm(O/H)}+12$ in the ISM, with varying dilution factor $f$ (represented by different markers on the curves), for the four scenarios defined in Sect.~\ref{Sec:IMF} (IMFs are represented by dashed or solid curves, mass-cuts are separated by column), and for only three of the six metallicities we investigate (Pop~III, cyan; EMP, green; and solar, grey). We have added data from observations in various galaxies, which we obtained from \citet{Marques2023}. Table~\ref{tab:all_ratios} presents our results in tabular form and includes the other three metallicities not shown on Fig.~\ref{fig:ratios}, for which we performed the same simulations (see Sect.~\ref{Sec:diffZ} for more details about these models). For each of the four scenarios, and each of the six metallicities, we give the range of values of $f$ for which $\log{({\rm O}/{\rm H})}+12$ fits within the fiducial confidence interval of \citet{Cameron2023}, if such a range exists. For each subsequent column, we provide the range of values predicted for that abundance ratio, for the acceptable dilution factors. If there is no acceptable dilution factor, we provide the range of values corresponding to dilution factors $1<f<1000$ (this is the case for the four scenarios at LMC and solar metallicities). If the two bounds of that range are within 0.01\,dex, we show only one value. Table~\ref{tab:obs_ratios} presents the observational data shown in Fig.~\ref{fig:ratios} in tabular form. In this section we discuss the three metallicities shown in Fig.~\ref{fig:ratios}, and we address the other three in the next section.

As the idea of enrichment by WR stars \citep[see e.g.][]{Kobayashi2023} has been proposed to explain the observed abundance ratios, we also show our results of WR models at solar metallicity in Fig.~\ref{fig:ratios}. Based on these results however, we rule out this possibility. Indeed, no matter the scenario and dilution factor, the (N/O) values are consistently too low ($\log{\rm(N/O)}\sim -1~{\rm to}~-0.6$), and/or the (O/H) values too high ($\log{\rm(O/H)}+12>8.7$) to match the observations. Since the only metallicities where both (O/H) and (N/O) ratios are able to match those measured in high-$z$ galaxies are $Z=0$ and $Z=10^{-5}$, we focus our discussion on these populations.

We obtain the best match to the observed (N/O) ratios in CEERS-1019 and GN-z11 for five of our populations of fast-rotating Pop~III and moderately-rotating EMP stars. Specifically, the `Top-heavy above remnant' at $Z=0$ ($\log{\rm(N/O)}\sim-0.22$) and `Salpeter above remnant' at $Z=10^{-5}$ (${\log{\rm(N/O)}\sim-0.23 \text{ -- } -0.24}$) populations fall within the observed $\log{\rm(N/O)}$ confidence interval of CEERS-1019. The `Top-heavy above CO core' at $Z=0$ and $Z=10^{-5}$ ($\log{\rm(N/O)}\sim-0.35$ and $-0.36$ respectively), as well as the `Top-heavy above remnant' at $Z=10^{-5}$ ($\log{\rm(N/O)}\sim-0.38$) populations reproduce that of GN-z11.

The two `Salpeter' scenarios at $Z=0$ do not fit within the observed confidence intervals for either CEERS-1019 or GN-z11, with $\log{\rm(N/O)}\sim-0.55~{\rm and}~-0.69$ for `Salpeter above remnant' and `Salpeter above CO core' respectively. These two populations are a better match to the observed (N/O) values in the Lynx arc \citep[$z=3.357$, $\log{\rm(N/O)}\sim-0.53$,][]{Fosbury2003} and SMACS2031 \citep[$z=3.5$, $\log{\rm(N/O)}\sim-0.66$,][]{Patricio2016}. 

The (C/O) ratios for the Pop~III and EMP populations are too high compared to the observational measurement of CEERS-1019 (they are a better match to the data from Mrk 996 and the Lynx arc) ,with $\log{\rm(C/O)}\sim-0.2~\text{--}~0$, except for the `Top-heavy above remnant' scenario at $Z=10^{-5}$ ($\log{\rm(C/O)}\sim-0.57$), whose (C/O) ratio is too low.

The impact of the IMF on the predicted (N/O) ratios differs among these metallicities: the `Top-heavy' scenarios predict values $0.23-0.34$\,dex larger than the `Salpeter' ones for the Pop~III populations, while for the EMP populations they predict values $0.15$\,dex larger (`above CO core') and $0.15$\,dex smaller (`above remnant') than the `Salpeter' scenarios. Apart from the latter case of the `above remnant' scenarios at $Z=10^{-5}$, this result is consistent with the fact that the major producers of nitrogen are the more massive stars. We attribute the different relative behaviours of the `above remnant' scenarios at $Z=10^{-5}$ to the very large quantity of oxygen ejected by the 120\,$M_\odot$ model at that metallicity when the PISN is considered. This reduces the (N/O) ratio of the `Top-heavy' scenario. The top-heavy scenarios predict smaller (C/O) ratios for the $Z=10^{-5}$ populations ($0.28-0.08$\,dex) and larger ratios for the $Z=0$ ones ($0.01-0.07$\,dex) than the Salpeter ones. While the result at $Z=10^{-5}$ can be explained by the fact that carbon is more abundantly produced than oxygen by the less massive stars, the one at $Z=0$ is more difficult to explain. It may be due to the incomplete computations of the models (especially the 120\,$M_\odot$ model which has not finished core He-burning). We discuss this case of the fast-rotating Pop~III 120\,$M_\odot$ model in Sect.~\ref{Sec:caveats}.

The impact of the mass-cut above which matter is ejected at the end of stars' lives is not straight-forward, although minor. When ejecting above the future remnant's mass coordinate, the mass coordinate of the mass-cut is larger than that of the CO core for the 20, 60, and 85\,$M_\odot$ stars (by 4\,$M_\odot$, see Sect.~\ref{Sec:remnants}), equal to that of the CO core for the 9\,$M_\odot$ stars (which become white dwarfs), and much smaller to that of the CO core for the 120\,$M_\odot$ stars (which end their lives in a PISN, with $M_{\rm rem}=0$). As a result, more matter is ejected from the highest-mass star, less from the intermediate-mass ones, and the same amount from the lowest-mass one. Broadly put, for the Pop~III populations, ejecting above the CO core lowers the (N/O) ratio (by $0.13~\text{--}~0.14$\,dex) compared to ejecting above the remnant. For the EMP populations, ejecting above the CO core lowers the (N/O) ratio (by $0.17$\,dex) for the `Salpeter' populations and raises it (by $0.02$\,dex) for the `Top-heavy' populations, compared to ejecting above the remnant. It raises the (C/O) ratios of populations at both metallicities (by $0.04-0.1$\,dex for $Z=0$, and by $0.01-0.38$\,dex for $Z=10^{-5}$).

The effect of the dilution factor $f$ is easiest to understand for the stellar populations at $Z=0$. Indeed, since their initial composition (and that of the unenriched ISM) contain only hydrogen and helium, dilution does not affect the (N/O) and (C/O) ratios. That is why the Pop~III curves in Fig.~\ref{fig:ratios} are horizontal lines, from right to left for increasing values of $f$. For the EMP populations, since their initial abundance of $^{16}$O is larger than those of $^{12}$C and $^{14}$N, their curves shift downwards for larger dilution factors.

We treat the dilution factor $f$ as a free parameter in our model, but comparing our results to the observed confidence interval of $\log{\rm(O/H)}+12$ can constrain its value. Using as constraint the fiducial (pink-shaded region in Fig.~\ref{fig:ratios}) interval of $7.62<\log{\rm(O/H)}+12<8$ from \citet{Cameron2023}, we obtain values in a general range of $15<f<159$. The specific ranges vary from $15<f<36$ (`Salpeter above remnant' at $Z=10^{-5}$) to $67<f<159$ (`Salpeter above CO core' at $Z=0$). Using the fiducial value of $\log{\rm(O/H)}+12=7.82$ from \citet{Cameron2023}, we find, for the three populations (`Top-heavy above remnant' and `Top-heavy above CO core' at $Z=10^{-5}$, and `Top-heavy above CO core' at $Z=0$) for which the (N/O) ratios can match the confidence interval of GN-z11, $67<f<100$. We note that, most notably in the populations with a Salpeter IMF, the $Z=0$ populations require a larger dilution of their ejecta to match the observed $\log{\rm(O/H)}+12$ than the $Z=10^{-5}$ ones.

\subsubsection{ Other metallicities} \label{Sec:diffZ}

For the sake of completeness, we perform the same computations for populations of rotating massive stars at three other metallicities ($Z=4\times10^{-4}$ (IZw18), $Z=0.002$ (SMC), and $Z=0.006$ (LMC)). Because the resulting (N/O) ratios are consistently too small compared to the ones measured in high-$z$ galaxies, we do not show the results for these three metallicities in Fig.~\ref{fig:ratios} (we nevertheless give the obtained abundance ratios in Table~\ref{tab:all_ratios}).

For the populations at $Z=4\times10^{-4}$ and $Z=0.002$, there exist values of $f$ such that $\log{\rm(O/H)}+12$ fits within the fiducial range for GN-z11 \citep{Cameron2023}, although for SMC metallicity a large ($f\sim1000$) dilution factor is required to approach the low-end of that range. No matter the scenario however, these populations systematically yield a (N/O) that is too small (at best we obtain $\log{({\rm N}/{\rm O})}=-0.89$ for the `Salpeter above remnant' population at SMC metallicity). It is interesting to note that the populations at IZw18 metallicity yield similar (O/H) and (C/O) ratios as the EMP ones, and that the (O/H) ratios are attained with similar dilution factor values, but that they consistently predict (N/O) ratios that are 1\,dex or more smaller.

The populations at LMC metallicity, like the ones at solar metallicity, systematically yield too large (O/H) and too small (N/O) ratios. Even at the largest dilution factor, their $\log{\rm(O/H)}+12$ are larger than the upper bound of the possible range for GN-z11 by \citet{Cameron2023}. This is because, for $f=1000$, the ejecta is so diluted that the composition of the enriched ISM is almost that of the unenriched ISM. At LMC metallicity the unenriched ISM has $\log{\rm(O/H)}+12=8.32$ and at solar metallicity, $\log{\rm(O/H)}+12=8.70$. It is thus not possible for these populations to match measured ratios in high-$z$ galaxies: even if the (N/O) could be reproduced, the ISM would nevertheless be too rich in oxygen.

\subsubsection{Predictions of other ratios} \label{Sec:ratios_pred}

In this section, we make predictions of abundance ratios that could be measured in high redshift galaxies. While some of our fast-rotating Pop~III and moderately-rotating EMP populations are able to match the observed (N/O) and (C/O) number ratios, future observations should confirm (or refute) the predictions we make about other isotopes in order to validate (or reject) our proposed enrichment process. To that end, we have also computed $\log{\rm(He/H)}$ and $\log{\rm(^{12}C/^{13}C)}$ for the eight (four scenarios at the two metallicities of $Z=0$ and $Z=10^{-5}$) populations. We present their values for the best-fit dilution factor of each population, that is the value of $f$ for which $\log{({\rm O}/{\rm H})}+12$ is closest to the fiducial value for GN-z11 \citep[7.82,][]{Senchyna2023}.\\

The (He/H) ratios are remarkably similar for the eight sets of models, ranging from $\log{\rm(He/H)}=-1.059$ (`Salpeter above remnant' at $Z=10^{-5}$) to $\log{\rm(He/H)}=-1.077$ (`Salpeter above CO core' at $Z=0$). These values correspond to He:H ratios between 1:11.5 and 1:11.9 (compared to the primordial 1:12 ratio). They are probably too similar for a measurement of this particular ratio to be able to distinguish between the scenarios. However, at solar metallicity we find $\log{\rm(He/H)}\sim-1.03$, which corresponds to a He:H ratio of 1:10.7. Observations should be able to discriminate between the ratios we predict at zero to extremely low metallicity, and those at solar metallicity.

The ($^{12}$C/$^{13}$C) ratios on the other hand are less degenerate. They are larger for the Pop~III populations, larger for scenarios using a Salpeter IMF, and larger for scenarios ejecting all matter above the CO core. For the Pop~III populations, we find $\log{\rm(^{12}C/^{13}C)}\sim2.18-2.42$ (`Salpeter above remnant' - `Salpeter above CO core') and $\log{\rm(^{12}C/^{13}C)}\sim1.71-1.84$ (Top-heavy, same order as above). For the EMP populations, we obtain $\log{\rm(^{12}C/^{13}C)}\sim1.42-1.68$ (Salpeter) and $\log{\rm(^{12}C/^{13}C)}\sim1.35-1.65$ (Top-heavy). These are different enough to discriminate between the $Z=0$ and $Z=10^{-5}$ models.

\section{Discussion}\label{Sec:Discussion}

\subsection{End points of stellar computations} \label{Sec:caveats}
We note that the 120\,$M_\odot$ fast-rotating Pop~III model has not depleted the helium in its core, and this may have an impact on the results of this study. The core He-burning phase is crucial for the evolution and nucleosynthesis of a star. Unfortunately, handling computations for extreme cases like the 120\,$M_\odot$ model is challenging due to the high initial rotation and very low metallicity. We are aware that the amount of ejected carbon would decrease if computations were able to reach the end of core He-burning. Additionally, nitrogen and oxygen are expected to increase during this process. Therefore, we anticipate an increase in the (N/O) ratio and a decrease in the (C/O) ratio, most notably for the populations computed with a top-heavy IMF.

Additionally, we have checked the impact on the predicted ratios of N/O and C/O for the populations if we integrate the chemical abundances in stars at their last computed timestep as compared to our current approach of homogeneously taking the yields at the end of core C-burning. While the predicted ratios change, the difference is not very substantial and the scenarios can still explain the observed ratios.

\subsection{Moderately versus fast-rotating massive stars} \label{Sec:fast}

We select a set of moderately-rotating Pop~III models \citep[taken from][]{Murphy2021a} with masses 9, 20, 60, 85, and 120\,$M_\odot$ until the end of He-burning to compare with our fast-rotating models. Accounting for dilution, we obtain for the fast-rotating models in the `Top-heavy above remnant' scenario $\log{\rm (N/O)}=-0.22$, $\log{\rm (C/O)}=-0.07$ and $\log{\rm (N/O)}+12=7.82$ (at f = 100). For the moderately-rotating Pop~III models, we obtain -2.93, -0.295, and 7.74 respectively, far below the observed values of (N/O) for GN-z11 and CEERS-1019 (see Fig.~\ref{fig:ratios}). We see similar differences when looking at the other three scenarios of moderately-rotating Pop~III models. This is due to the stronger transport of $^{12}$C from the core to the H-burning shell in the case of fast-rotating Pop~III models, as discussed in Sect.~\ref{Sec:Transport}. A detailed analysis on the transport mechanisms of fast rotating massive stars is described in the upcoming work by \citet{Tsiatsiou2024}.

\subsection{Ejection above remnant versus CO core} \label{Sec:remnant_vs_CO}

In this work, we follow two different approaches to determine the mass-cut above which elements are ejected at the end of a star's life. One follows the results of \citet{Patton2020} and \citet{Farmer2019} and the other one simply ejects all matter above the mass coordinate of the CO core. The latter is very simplistic and most certainly inaccurate. However, the former may also present inaccuracies, the most impactful of which being the prediction of whether the remnant will be a NS or a BH. That prediction can dramatically change the value of the mass cut. Rotation was not taken into account by \citet{Patton2020}, although it has been shown to support neutrino-driven explosions \citep[see e.g.][]{Muller2020,Buellet2023}. For most stars it is very clear what type of remnant will result from their supernova. However, the 20\,$M_\odot$ stars lie close to the border between implosion and explosion. For example, the 20\,$M_\odot$ EMP star has, at the end of core He-burning, a CO core mass $M_{\rm CO}=2.98\,M_\odot$ and central carbon mass fraction $x(^{12}{\rm C})_c=0.26$, so Fig.~3 of \citet{Patton2020} predicts it to implode into a black hole with mass $M_{\rm BH}=6.98\,M_\odot$. If however it had $x(^{12}{\rm C})_c=0.28$, then an explosion would be predicted, leaving behind a much less massive ($M_{\rm NS}=1.64\,M_\odot$) neutron star. As a result it would eject 5.34\,$M_\odot$ more material, and this material would be metal-rich because it is located in the inner regions of the star. A similar reasoning can be applied for the high-mass range of our models, where the difference between a PISN and a pulsational PISN results in 30\,$M_\odot$ of matter being ejected or locked into a black hole.

While we believe our results are robust to these uncertainties on the type of remnant, as shown by the relatively modest variations between the left and right columns in Fig.~\ref{fig:ratios}, this caveat remains one to keep in mind.

\subsection{Impact of intermediate mass stars}

In this work, we study the chemical impact of stars more massive than 9\,$M_\odot$. While this makes sense for Pop~III stars that are expected to be massive and form from a top-heavy IMF \citep{Bromm2004,Greif2011,Stacy2013,Susa2014,Wollenberg2020}, intermediate mass (3-8\,$M_\odot$) stars should form at EMP and higher metallicities. These stars can also greatly enrich the surrounding ISM during their asymptotic giant branch (AGB) phase. As a result, our models provide a snapshot picture of the chemical enrichment in young galaxies ($\sim10$\,Myr old), before the intermediate mass stars would have evolved to the AGB phase. Since we consider high-$z$ galaxies, which may indeed be quite young, we think that limiting our study to massive stars is justified. We nevertheless keep in mind that the galaxies will evolve and so will their chemical composition.

\subsection{Comparison with previous studies} \label{Sec:Comparison}

The precise measurement of chemical enrichment in high-$z$ galaxies has only recently been made possible following the launch of JWST. In this section, we compare our results with other works published on this subject. 

\citet{Kobayashi2023} introduced a dual starburst model to show the high nitrogen abundances in GN-z11. They proposed that star formation began around a redshift of $z \sim 16.7$, lasting 100 million years, followed by an equally long quiescent phase. The model highlights WR stars during the second starburst, at metallicities [Fe/H] = 0, -1, -2, and -3, as sources for elevated C/O and N/O ratios. Utilising this model, they report $\log{\rm (N/O)}=0.246$, in line with observed data when calibrated against $\log{\rm (O/H)}+12$. Our models at solar metallicity (${\rm [Fe/H]} = 0$) employ a simpler approach using stellar yields and an initial mass function. Despite this, our models at solar metallicity, primarily based on the ejecta from WR stars, do not replicate \citet{Kobayashi2023}'s results. The latter account for star formation, gas inflows and outflows in their one-zone model which includes yields of asymptotic giant branch stars, super-AGB stars, and core-collapse supernovae. Whereas we explore the problem from the side of stellar physics and use a parameterised prescription for the chemical enrichment of the ISM. This highlights the differences between our yield-focused methodology and their more comprehensive galactic evolution model.

Comparing the results from \citet{Nandal2024}. and our study, both using the Geneva stellar evolution code, we find similarities and differences in elemental ratios. Since the ejection mechanism of SMSs is unknown, \citet{Nandal2024} assume the ejection of chemical species at three different mass cuts ranging from near the surface to the CO core, whereas models explored in our work are assumed to eject mass either above the remnant or CO core. They explore a mass range of 500-9000 M$_\odot$ whereas this work focuses on the standard massive stars ranging between 9-120 M$_\odot$. \citet{Nandal2024} show that Pop~III stars between masses \(1900-9000 \, M_{\odot}\) have a \(\log\) (N/O) ratio of -0.37, a value we also find in our study for fast-rotating massive Pop~III stars. This shows agreement in N/O ratios across different mass ranges. In terms of C/O ratios, \citet{Nandal2024} provide a value of -0.65. In our research, fast-rotating Pop~III stars give a \(\log{\rm (C/O)} = -0.2~\text{--}-0.06\). O/H ratios from \citet{Nandal2024} and our work (\(\log{\rm (O/H)} + 12 \sim 7.75\)) are similar. \citet{Nandal2024} predict environments rich in \(^4\)He, with a mass fraction of 0.55 to 0.65, while our study predicts \(\log{\rm(He/H)}\) ranging from -1.077 to -1.059. These comparisons reveal alignment in N/O, C/O and O/H ratios, but differences in other ratios like He/H.

\citet{Nagele2023} proposed that supermassive stars with masses from $10^3$ to $10^5\,M_\odot$ at $Z=0.1\,Z_\odot$ are necessary to the super-solar nitrogen levels in GN-z11, with $\log{\rm (N/O)}$ between -0.9 and 1.25, and $\log{\rm (C/O)}$ between -1.7 and 1.25. Our work indicates that fast-rotating massive Pop~III stars can replicate GN-z11’s observed ratios without invoking SMSs. This suggests that fast-rotating massive stars, in addition to SMSs, can produce these observed ratios, aligning with data for GN-z11 and CEERS-1019. However, we concur with \citet{Nagele2023} that standard metal-enriched and moderately-rotating Pop~III models fail to reproduce the observed abundances, highlighting the need for alternative scenarios in explaining GN-z11’s chemical composition.

\citet{Charbonnel2023} explored GN-z11's chemical composition using the runaway collision scenario put forward by \citet{Gieles2018}, focusing on a $10^4\,M_{\odot}$ SMS with [Fe/H] between -1.76 and -0.78. They highlighted a fully convective SMS that grows through collisions. In contrast, our work employs a massive Pop~III star formation model, where fast rotation leads to efficient transport of elements. Our Pop~III models achieve higher values of N/O and C/O, with O/H matching observed values at dilution factors of $20~\text{--}~100$.

From their study of 70 galaxies containing star forming regions at $z = 4~\text{--}~10$, \citet{Isobe2023} classify GN-z11, CEERS-1019 and GLASS-150008 as N-rich galaxies. They use chemical evolution models rich in CNO elements involving WR, supermassive stars and tidal disruption events to reproduce the N/O and C/O ratios. Their values of $\log{\rm (N/O)}>0.28$ and $\log{\rm (C/O)}<-1.04$ for CEERS-1019 and $\log{\rm (N/O)}>-0.36$ and $\log{\rm (C/O)}>-1.01$ for GN-z11 are similar to ours for the fast-rotating Pop~III models. \citet{Isobe2023} state the need of frequent direct collapse to obtain such ratios whereas our ratios are obtained from models that undergo core collapse at the end of their evolution. 

\citet{Vink2023} highlighted the potential of very massive stars (100-1000\,$M_\odot$) in nitrogen enrichment, focusing on their advantages over SMSs and WR stars. While their study did not detail specific models, it underscored the significance of VMSs in galactic nucleosynthesis. Our research complements this by computing models for massive stars up to 120\,$M_{\odot}$. These models, demonstrating characteristics akin to VMSs, can potentially reproduce observed galactic chemical abundances. 

\citet{Marques2023} analysed CEERS-1019 and GN-z11, noting supersolar ratio of $\log{\rm (N/O)}$ in CEERS-1019. To explain these abundances, they considered two scenarios: enrichment from WR stars and nucleosynthesis from supermassive stars with masses over 1000\,$M_\odot$. In contrast, the abundance ratios from our Pop~III stars with $\upsilon_{\mathrm{ini}}/\upsilon_{\mathrm{crit}} = 0.7$ can match those in GN-z11 and CEERS-1019 without requiring SMSs. Additionally, our WR models (at solar metallicity) do not reproduce the observed ratios and therefore do not match the results from \citet{Marques2023}.

\section{Conclusion}\label{Sec:Conclusion}

In this study, we have analysed the chemical compositions of high-redshift galaxies, focusing specifically on the nitrogen-enhanced GN-z11 and CEERS-1019. Using {\sc Genec} stellar models, our research has provided insightful findings into the chemical processes in early galaxies. We have computed the ejecta from massive stars, ranging from 9 to 120\,$M_\odot$, across various metallicities, and integrated these findings with observations from high-redshift galaxies. Our results reveal the intricate role of stellar mass, rotation, and metallicity in shaping the chemical landscape of the early universe. Key findings of our research include:
\begin{itemize}
  \item Populations of fast-rotating ($\upsilon_{\mathrm{ini}}/\upsilon_{\mathrm{crit}} = 0.7$) Population III stars following a top-heavy IMF, and moderately-rotating $Z=10^{-5}$ populations can match the abundance ratios of (N/O) and (O/H) in either GN-z11 or CEERS-1019, with values $\log{\rm (N/O)} = -0.35~\text{--}~-0.22$ ($Z=0$), $\log{\rm (N/O)} = -0.38~\text{--}~-0.23$ ($Z=10^{-5}$), and $\log{\rm (O/H)} = + 12 = 7.82$ at dilution factors of $f \sim 75~\text{--}~100$ ($Z=0$) and $f \sim 22~\text{--}~71$ ($Z=10^{-5}$).
  \item Models at all other metallicities ($Z=4\times 10^{-4}, 0.002, 0.006, 0.014$) fail to match these observed abundance ratios, thereby underscoring the unique nature of rotating EMP and fast-rotating Population III stars in early galaxy evolution.
  \item Predictions for other abundance ratios, such as (He/H) and ($^{12}$C/$^{13}$C), provide crucial benchmarks for future observations. The predicted ranges include $\log{\rm(He/H)}$ from -1.077 to -1.059 and $\log{\rm(^{12}C/^{13}C)}$ from 1.35 to 2.42.
  \item Our models indicate that different stellar masses significantly impact the production of key elements. Specifically, lower mass stars (9-20\,$M_\odot$) are predominant producers of $^{16}$O, 60\,$M_\odot$ models are the primary producers of $^{14}$N, whereas higher mass stars (85-120\,$M_\odot$), especially those undergoing pair-instability supernovae, are crucial for the synthesis of $^{12}$C. This variation in elemental production across different masses underlines the complex interplay between stellar mass and nucleosynthetic processes.
\end{itemize}

Our work contributes to a deeper understanding of the chemical evolution of early galaxies, emphasising the pivotal role of massive stars, particularly those with fast rotation rates. The next steps in this research could include exploring models that incorporate the impact of internal magnetic fields on stellar evolution and nucleosynthesis. Additionally, integrating the yields from our stellar models into comprehensive galactic chemical evolution models would provide a more holistic view of galactic formation and evolution in the early universe. These future endeavours will further enhance our understanding of the intricate processes shaping the cosmos.

\section*{Acknowledgements}
The three authors have equally contributed to this research and D.N., Y.S, and S.T. are all to be considered as first authors of this work (the names were placed in alphabetical order).
We would like to thank the referee for their constructive comments and suggestions that helped to improve the paper. D.N., Y.S., and S.T. have received funding from the European Research Council (ERC) under the European Union's Horizon 2020 research and innovation programme (grant agreement No. 833925, project STAREX). The three authors would like to express their gratitude to their supervisors Prof. Georges Meynet, Dr. Sylvia Ekstr\"{o}m, and Dr. Cyril Georgy for their supervision and support during this project.

\bibliographystyle{aa}
\bibliography{main}

\end{document}